\documentclass[twoside,twocolumn,9pt]{article}
\usepackage{extsizes}
\usepackage[super,sort&compress,comma]{natbib} 
\usepackage[version=3]{mhchem}
\usepackage[left=1.5cm, right=1.5cm, top=1.785cm, bottom=2.0cm]{geometry}
\usepackage{balance}
\usepackage{widetext}
\usepackage{times,mathptmx}
\usepackage{sectsty}
\usepackage{graphicx} 
\usepackage{lastpage}
\usepackage[format=plain,justification=raggedright,singlelinecheck=false,font={stretch=1.125,small,sf},labelfont=bf,labelsep=space]{caption}
\usepackage{float}
\usepackage{fancyhdr}
\usepackage{fnpos}
\usepackage[english]{babel}
\usepackage{array}
\usepackage{droidsans}
\usepackage{charter}
\usepackage[T1]{fontenc}
\usepackage[usenames,dvipsnames]{xcolor}
\usepackage{setspace}
\usepackage[compact]{titlesec}
\definecolor{cream}{RGB}{222,217,201}

\begin{document}

\pagestyle{fancy}
\thispagestyle{plain}
\fancypagestyle{plain}{

%%%HEADER%%%
%\fancyhead[C]{\includegraphics[width=18.5cm]{header_bar}}
%\fancyhead[L]{\hspace{0cm}\vspace{1.5cm}\includegraphics[height=30pt]{journal_name}}
%\fancyhead[R]{\hspace{0cm}\vspace{1.7cm}\includegraphics[height=55pt]{RSC_LOGO_CMYK}}
\renewcommand{\headrulewidth}{0pt}
}
%%%END OF HEADER%%%

%%%PAGE SETUP - Please do not change any commands within this section%%%
\makeFNbottom
\makeatletter
\renewcommand\LARGE{\@setfontsize\LARGE{15pt}{17}}
\renewcommand\Large{\@setfontsize\Large{12pt}{14}}
\renewcommand\large{\@setfontsize\large{10pt}{12}}
\renewcommand\footnotesize{\@setfontsize\footnotesize{7pt}{10}}
\makeatother

\renewcommand{\thefootnote}{\fnsymbol{footnote}}
\renewcommand\footnoterule{\vspace*{1pt}% 
\color{cream}\hrule width 3.5in height 0.4pt \color{black}\vspace*{5pt}} 
\setcounter{secnumdepth}{5}

\makeatletter 
\renewcommand\@biblabel[1]{#1}            
\renewcommand\@makefntext[1]% 
{\noindent\makebox[0pt][r]{\@thefnmark\,}#1}
\makeatother 
\renewcommand{\figurename}{\small{Fig.}~}
\sectionfont{\sffamily\Large}
\subsectionfont{\normalsize}
\subsubsectionfont{\bf}
\setstretch{1.125} %In particular, please do not alter this line.
\setlength{\skip\footins}{0.8cm}
\setlength{\footnotesep}{0.25cm}
\setlength{\jot}{10pt}
\titlespacing*{\section}{0pt}{4pt}{4pt}
\titlespacing*{\subsection}{0pt}{15pt}{1pt}
%%%END OF PAGE SETUP%%%

%%%FOOTER%%%
\fancyfoot{}
%\fancyfoot[LO,RE]{\vspace{-7.1pt}\includegraphics[height=9pt]{LF}}
%\fancyfoot[CO]{\vspace{-7.1pt}\hspace{13.2cm}\includegraphics{RF}}
%\fancyfoot[CE]{\vspace{-7.2pt}\hspace{-14.2cm}\includegraphics{RF}}
\fancyfoot[RO]{\footnotesize{\sffamily{1--\pageref{LastPage} ~\textbar  \hspace{2pt}\thepage}}}
\fancyfoot[LE]{\footnotesize{\sffamily{\thepage~\textbar\hspace{3.45cm} 1--\pageref{LastPage}}}}
\fancyhead{}
\renewcommand{\headrulewidth}{0pt} 
\renewcommand{\footrulewidth}{0pt}
\setlength{\arrayrulewidth}{1pt}
\setlength{\columnsep}{6.5mm}
\setlength\bibsep{1pt}
%%%END OF FOOTER%%%

%%%FIGURE SETUP - please do not change any commands within this section%%%
\makeatletter 
\newlength{\figrulesep} 
\setlength{\figrulesep}{0.5\textfloatsep} 

\newcommand{\topfigrule}{\vspace*{-1pt}% 
\noindent{\color{cream}\rule[-\figrulesep]{\columnwidth}{1.5pt}} }

\newcommand{\botfigrule}{\vspace*{-2pt}% 
\noindent{\color{cream}\rule[\figrulesep]{\columnwidth}{1.5pt}} }

\newcommand{\dblfigrule}{\vspace*{-1pt}% 
\noindent{\color{cream}\rule[-\figrulesep]{\textwidth}{1.5pt}} }

\makeatother
%%%END OF FIGURE SETUP%%%

%%%TITLE, AUTHORS AND ABSTRACT%%%
\twocolumn[
  \begin{@twocolumnfalse}
\vspace{3cm}
\sffamily
\begin{tabular}{m{4.5cm} p{13.5cm} }

 & \noindent\LARGE{\textbf{Ergodicity breaking of iron displacement in heme proteins$^\dag$}} \\
\vspace{0.3cm} & \vspace{0.3cm} \\

 & \noindent\large{Salman Seyedi\textit{$^{a}$} and Dmitry V.\ Matyushov$^{\ast}$\textit{$^{b}$} } \\

& \noindent\normalsize{We present a model of the dynamical transition of atomic displacements in proteins. Increased mean-square displacement at higher temperatures is caused by softening of the vibrational force constant by electrostatic and van der Waals forces from the protein-water thermal bath. Vibrational softening passes through a nonergodic dynamical transition when the relaxation time of the force-force correlation function enters,  with increasing temperature, the instrumental observation window. Two crossover temperatures are identified. The lower crossover, presently connected to the glass transition, is related to the dynamical unfreezing of rotations of water molecules within nanodomains polarized by charged surface residues of the protein. The higher crossover temperature, usually assigned to the dynamical transition, marks the onset of water translations. All crossovers are ergodicity breaking transitions depending on the corresponding observation windows. Allowing stretched exponential relaxation of the protein-water thermal bath significantly improves the theory-experiment agreement when applied to solid protein samples studied by M{\"o}ssbauer spectroscopy. } \\

\end{tabular}

 \end{@twocolumnfalse} \vspace{0.6cm}

  ]
%%%END OF TITLE, AUTHORS AND ABSTRACT%%%

%%%FONT SETUP - please do not change any commands within this section
\renewcommand*\rmdefault{bch}\normalfont\upshape
\rmfamily
\section*{}
\vspace{-1cm}

%%%FOOTNOTES%%%

\footnotetext{\textit{$^{a}$Department of Physics, Arizona State University, PO Box 871504, Tempe, Arizona 85287. }}
\footnotetext{\textit{$^{b}$Department of Physics and School of Molecular Sciences, Arizona State University, PO Box 871504, Tempe, Arizona 85287; E-mail: dmitrym@asu.edu }}

%Please use \dag to cite the ESI in the main text of the article.
%If you article does not have ESI please remove the the \dag symbol from the title and the footnotetext below.
\footnotetext{\dag~Electronic Supplementary Information (ESI) available: [details of any supplementary information available should be included here]. See DOI: 10.1039/b000000x/}
%additional addresses can be cited as above using the lower-case letters, c, d, e... If all authors are from the same address, no letter is required

%  \footnotetext{\ddag~Additional footnotes to the title and authors can be included \emph{e.g.}\ `Present address:' or `These authors contributed equally to this work' as above using the symbols: \ddag, \textsection, and \P. Please place the appropriate symbol next to the author's name and include a \texttt{\textbackslash footnotetext} entry in the the correct place in the list.}

%%%END OF FOOTNOTES%%%

%%%MAIN TEXT%%%%
\section{Introduction}
Atomic displacements in proteins are viewed as a gauge of the overall flexibility of macromolecules.  Displacements of the hydrogen atoms are reported by neutron scattering,\cite{Lovesey} and mean-square displacements (B-factors) of all atoms are known from X-ray crystallography. Neutron scattering reports ensemble averages of scattering from many hydrogen atoms of a single protein.\cite{Lovesey,Gabel:02,Smith:91}  In contrast, M{\"o}ssbauer spectroscopy often probes the displacements of a single atom in the protein,\cite{Frauenfelder:88,Parak:03} which is the heme iron in this study focused on cytochrome \textit{c} (Cyt-c) and myoglobin proteins. 

The temperature dependence of atomic displacements from both neutron scattering and M{\"o}ssbauer spectroscopy shows a number of crossovers. They are marked by changes in the slope of atomic mean-square displacement vs temperature,\cite{Parak:71,DosterNature:89} deviating from expectations from the fluctuation-dissipation theorem.\cite{Kubo:66,Achterhold:02} This problem has attracted significant attention in the literature.\cite{Gabel:02,Frauenfelder:09,Parak:03,DosterBBA:10,Ngai:2017kj,2016NCimC..39..305S}  The accumulation of the data over several decades of studies, combined with their recent refinements through the comparison of the results obtained on spectrometers with different resolution,\cite{Magazu:2011kz,Capaccioli:2012jc,Ngai:2013ic,MAGAZU:2017bn} have lead to a convergent phenomenological picture.  

\begin{figure}
\includegraphics*[width=6cm]{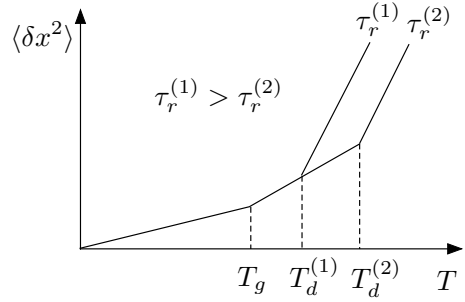}
\caption{Schematic representation of two crossovers in the temperature dependence of the mean-square fluctuation (MSF) $\langle \delta x^2\rangle$. The lower crossover, $T_g$, is independent of the instrumental resolution window and corresponds to the glass transition of the protein-water interface. The upper crossover (dynamical transition), $T_d$, does depend on the observation window and is related to the entrance of the relaxation time of the force acting on the coarse-grained unit (residue, cofactor, etc) into the resolution window of the experiment. The temperature $T_d$ shifts to the lower value when the observation time is increased.  
}
\label{fig:1}  
\end{figure}

Two low-temperature crossovers are now identified (Fig.\ \ref{fig:1}). The higher-temperature crossover $T_d$, originally assigned to the protein dynamical transition,\cite{DosterNature:89,DosterBBA:10} depends on the observation window of the spectrometer\cite{Frauenfelder:09,Magazu:2011kz,Capaccioli:2012jc,Ngai:2013ic} and shifts to lower temperatures when the resolution is increased  (a longer observation time $\tau_r$ in Fig.\ \ref{fig:1}). The lower crossover temperature, $T_g\simeq 170-180$ K, is independent of the observation window (in the range of resolution windows available to spectroscopy) and is assigned to the glass transition of the protein hydration shell.\cite{DosterNature:89,Ngai:2013ic,DosterBBA:10,Capaccioli:2012jc} 

All motions, rotations and translations, in the hydration shell (except for cage rattling) terminate at the lower temperature $T_g$. While this interpretation is consistent with the basic phenomenology of glass science, it does not address the question of how the structure and  dynamics of the hydration shell affect atoms inside the protein, the heme iron for M{\"o}ssbauer spectroscopy. The basic question here is whether the observations can be fully related to stiffening of the hydration shell at lower temperatures, thus reducing elastic deformations of the protein,\cite{Hong:2013bk} or there are some long-range forces acting on the heme, which are reduced in their fluctuations when the hydration shell dynamically freezes. It is possible that no simple answer to this question can be obtained in the case of neutron scattering since there are several classes of motions of protein hydrogens: cage rattling, methyl rotations, and  jumps between cages.\cite{Hong:2011qf} To avoid these complications, we focus here on a single heavy atom, heme iron, probed by M{\"o}ssbauer spectroscopy on the resolution time $\tau_r=142$ ns.     

The question addressed here is what are the physical mechanisms propagating fluctuations of the protein-water interface to an internal atom within the protein.\cite{Hong:2012dsa,Wood:2012bf} This question, also relevant to how enzymes work,\cite{DMjpcm:15} was addressed by the electro-elastic model of the protein,\cite{DMpre:11,DMjcp2:12} where both the effect of the viscoelastic deformation and the effect of the long-range forces acting on the heme iron were considered. The main conclusion of that theoretical work was the recognition of the two-step nature of the crossover in the mean-square fluctuation (MSF) of the heme iron. The low-temperature crossover, $T_g\simeq 170-180$ K, was assigned to an enhancement of viscoelastic deformations above the glass transition of the protein-water interface.\cite{DMpre:11} The increment in the MSF at $T_g$ was, however, insignificant, as  confirmed below based on new molecular dynamics (MD) simulations. It was, therefore, concluded that altering elastic stiffening is not sufficient to describe the rise of the MSF above $T_d$ and long-range forces need to be involved. 

The iron MSF significantly increases when electrostatic forces acting on the iron are included.\cite{DMpre:11} The dynamical transition and the corresponding enhancement of the MSF are promoted by ergodicity breaking when the longest relaxation time crosses the instrumental time. The equation for the MSF resulting from this perspective involves the MSF from local vibrations of the heme $\langle \delta x^2\rangle_\text{vib}$ and the global softening of the heme motions through the long-ranged forces acting on it. This second component enters the denominator of Eq.\ \eqref{eq:1} through the variance of the force acting on heme's iron $\langle \delta F^2\rangle_r$    

\begin{equation}
\langle \delta x^2\rangle_r = \frac{\langle \delta x^2\rangle_\text{vib}}{1-\beta^2\langle \delta F^2\rangle_r\langle \delta x^2\rangle_\text{vib} }
\label{eq:1}  
\end{equation}

The subscript ``r'' in the angular brackets, $\langle\dots\rangle_r$, indicates that the average is constrained by the observation window $\tau_r$. Correspondingly, the fluctuations of the long-range forces are mostly frozen at low temperatures when $\langle \delta F^2\rangle_r$ is low, yielding $\langle \delta x^2\rangle \simeq \langle \delta x^2\rangle_\text{vib}$. Since the relaxation time of the long-range forces $\tau(T)$ depends on temperature according to the Arrhenius law, it shortens with increasing temperature, ultimately reaching the point\cite{DMpre:11,Ngai:2011jk,DMjpcb:11,Ngai:2017kj} $\tau_r\simeq\tau(T_d)$, at which the high-temperature crossover occurs. Fluctuations of the long-range forces become dynamically unfrozen at this temperature, leading to an increase of both  $\langle \delta F^2\rangle_r$ and $\langle \delta x^2\rangle_r$. 

In the present paper, we present new extensive simulations of Cyt-c in solution at different temperatures. The goal is to assert the role of long-range forces in achieving the vibrational softening of atomic displacements at high temperatures (Eq.\ \eqref{eq:1}). We consider the entire heme as a separate unit experiencing the force from the surrounding thermal bath. This coarse graining allows us to focus on the long-time relaxation of the force-force correlation function relevant for the long observation time, $\tau_r=142$ ns, of the M{\"o}ssbauer experiment. We find that the longest relaxation time $\tau(T)$ follows the Arrhenius law with the activation barrier characteristic of a secondary relaxation process ($\beta$-relaxation of glass science\cite{Ediger:96}). We therefore support the proposal advanced by Frauenfelder and co-workers\cite{Frauenfelder:09,Fenimore:2013eo} that the higher-temperature crossover is caused by ergodicity breaking when the relaxation time of the secondary process characterizing the protein-water interface enters the experimental observation window. This relaxation process effects the heme iron through the combination of non-polar (van der Waals) and polar (electrostatic) forces. 

Our focus on the protein in solution has a limited applicability to experiments done with solid samples. Nevertheless, computer simulations produce results close to observations for the reduced state of Cyt-c. The length of simulations is also insufficient to sample the dynamics on the time-scale of $\tau_r=142$ ns. In addition, the solution setup does not reproduce highly stretched dynamics observed in protein powders.\cite{Hong:2011qf,Khodadadi:2015jp,Ngai:2017kj} We find that the agreement between theory and  experiment\cite{Parak:05} is much improved when stretched exponential dynamics from dielectric spectroscopy\cite{Nakanishi:2014gy} are used in our model.   

Despite limitations of our simulations in application to experimental data, there is one significant advantage of the solution setup. Experiments done with solid samples cannot claim that the observed phenomenology directly applies to solutions. The similarity between our simulations and such experiments gives credit to the idea that dynamical transition, caused by ergodicity breaking, is a general phenomenon relevant to physiological conditions. From a more fundamental perspective, ergodicity breaking is broadly applicable to enzymetic activity at physiological conditions and is described by a formalism carrying significant similarities with the problem of dynamical transition of atomic displacements.\cite{DMjpcm:15} 

\section{Formalism}
The standard definition adopted for the fraction of recoiless absorption of the $\gamma$-photon in M{\"o}ssbauer spectroscopy is through the average
\begin{equation}
f(k) = \left|\left\langle e^{ikx}\right\rangle_\text{r} \right|^2  .
\label{eq:3}
\end{equation}
The average $\langle\dots\rangle_r$ is over the statistical configurations of the system accessible on a given time resolution of the experiment specified through the observation (resolution) time $\tau_r$. Further, $k$ is the wavevector aligned with the $x$-axis of the laboratory frame and $x$ is the displacement of the heme iron. 

The average over the stochastic variable of iron displacement $x$ can be represented by an ensemble average with the free energy $F_r(x)$
\begin{equation}
\left\langle e^{ikx}\right\rangle_\text{r} = \int dx  e^{ik x -\beta F_r(x)} ,
\label{eq:4}  
\end{equation}
where $\beta=1/(k_\text{B}T)$ is the inverse temperature. 

The free energy $F_r(x)$ is distinct from the usual thermodynamic free energy in two regards. First, it is a partial free energy corresponding to the reversible work performed by all degrees freedom of the system at a fixed displacement $x$. Therefore, $F_r(x)$ is analogous to the Landau functional of the thermodynamic order parameter.\cite{Landau5} There is another distinction of $F_r(x)$ from the thermodynamic free energy specified by the subscript ``r''. This free energy is defined by sampling the constrained part of the phase space $\Gamma_r$ which can be accessed on the resolution time $\tau_r$. The definition of $F_r(x)$ should thus include two constraints: (i) a fixed value $x$ and (ii) a restricted phase space available to the system. Both constraints are mathematically realized by the following equation\cite{Palmer:82,Crisanti:2003wf,DMjpcm:15}
\begin{equation}
e^{-\beta F_r(x)} = \int_{\Gamma_r}  d\Gamma \delta\left(x - \mathbf{\hat x}\cdot\mathbf{q}\right) e^{-\beta H} .
\label{eq:5}  
\end{equation}
Here, $\mathbf{\hat x}$ is the unit vector along the $x$-axis and $\mathbf{q}$ is the iron's displacement vector. The restriction of the phase space is realized as a dynamical constraint on the frequencies over which the correlation functions appearing in the response functions are integrated.\cite{DMjpcm:15} A simple cutoff, $\omega>\omega_r=\tau_r^{-1}$, is used in the statistical averages below.        

We will next consider the displacement of the iron as composed of the displacement of the heme's center of mass and the normal-mode vibrations relative to the center of mass. The Hamiltonian in Eq.\ \eqref{eq:5} can therefore be separated into a linear term involving the external force $\mathbf{F}$ acting on the heme from the protein-water thermal bath and the Hamiltonian $H_\text{vib}$ of intra-heme vibrations 
\begin{equation}
H(\mathbf{q}) = H(0) - \mathbf{q}\cdot\mathbf{F} + H_\text{vib} .  
\label{eq:6}  
\end{equation}
By expanding the iron's displacement $\mathbf{q}$ in the normal-mode vibrations $\mathbf{Q}_\alpha$, we can re-write the free energy $F_r(x)$ in the form
\begin{equation}
\begin{split}
e^{-\beta F_r(x)+\beta H(0)} = &\int d\mathbf{q} \delta\left(x - \mathbf{\hat x}\cdot\mathbf{q}\right) \langle e^{\beta \mathbf{q}\cdot\mathbf{F}}\rangle_B \\ 
 &\int \prod_\alpha d\mathbf{Q}_\alpha \delta \left(\mathbf{q}-\sum_\alpha \mathbf{\hat e}_\alpha \frac{Q_\alpha}{\sqrt{m}} \right) e^{-\beta H_\text{vib}}  ,
\end{split}  
\label{eq:7}  
\end{equation}
where $m$ is the mass of the iron atom. Further, the average $\langle\dots\rangle_B$ is over the fluctuations of the classical protein-water thermal bath which creates movements of the heme as a whole. It is reasonable to anticipate that these relatively large-scale fluctuations follow the Gaussian statistics with the force variance $\sigma_F^2=\langle (\delta \mathbf{F})^2\rangle$, $\delta \mathbf{F}=\mathbf{F}-\langle \mathbf{F}\rangle$. The average over such fluctuations in Eq.\ \eqref{eq:7} then becomes
\begin{equation}
  \langle e^{\beta \mathbf{q}\cdot\mathbf{F}}\rangle_B = e^{(\beta q  \sigma_F)^2/2} .
  \label{eq:8}
\end{equation}
In addition, the integral over the normal modes in Eq.\ \eqref{eq:7} is a Gaussian integral such that  
\begin{equation}
\int \prod_\alpha d\mathbf{Q}_\alpha \delta \left(\mathbf{q}-\frac{1}{\sqrt{m}}\sum_\alpha \mathbf{\hat e}_\alpha Q_\alpha \right) e^{-\beta H_\text{vib}}  = e^{-q^2/(2\sigma_\text{vib}^2)},
\label{eq:9}  
\end{equation}
where the variance due to intramolecular vibrations is
\begin{equation}
\sigma_\text{vib}^2 = \frac{\hbar}{6m}\sum_\alpha \hat e_\alpha^2 \frac{2\bar n_\alpha+1}{\omega_\alpha} . 
\label{eq:10}  
\end{equation}
Here, $\bar n_\alpha $ is the average occupation number of the normal mode $\alpha$ with the frequency $\omega_\alpha$. By substituting Eqs.\ \eqref{eq:8} and \eqref{eq:9} into Eq.\ \eqref{eq:7}, one obtains the harmonic free energy function\cite{DMpre:11} 
\begin{equation}
\beta F_r(x) = H(0) + \frac{x^2}{2\sigma^2} 
\label{eq:11}  
\end{equation}
with the variance 
\begin{equation}
\sigma^2 = \frac{\sigma_\text{vib}^2 }{1-(\beta\sigma_F\sigma_\text{vib})^2} .  
\end{equation}

The basic result of this derivation is straightforward: adding Gaussian fluctuations of the heme's center of mass to intramolecular vibrations of the heme leads to the softening of the force constant of the harmonic free energy $F(x)$.\cite{Hong:2012dsa} Combining this result with Eqs.\ \eqref{eq:3} and \eqref{eq:4}, one obtains the Gaussian form for the recoiless fraction 
\begin{equation}
f(k) = e^{-k^2\langle \delta x^2\rangle_r}
\label{eq:12}  
\end{equation}
with $\langle \delta x^2\rangle_r$ given by Eq.\ \eqref{eq:1} in which
$\langle \delta x^2\rangle_\text{vib}=\sigma_\text{vib}^2$. 
 
The subscript ``r'' in $\langle \delta F^2\rangle_r$ specifies that the average over the stochastic fluctuations of the force $\mathbf{F}$ acting on the heme from the thermal bath is understood in the spirit of the dynamically restricted average over a dynamically accessible subspace of the system $\Gamma_r$, as specified in Eq.\ \eqref{eq:5}. In practical terms, this implies that only frequencies greater than $\omega_r=\tau_r^{-1}$ can contribute to the observables. The effective variance can therefore be calculated as\cite{Frauenfelder:09,DMjpcm:15}     
\begin{equation}
\langle \delta F^2\rangle_r = \int_{\omega_r}^{\infty} (d\omega/\pi) C_F(\omega)  .
\label{eq:14}  
\end{equation}
Here, $C_F(\omega)$ is the Fourier transform of the time auto-correlation function
\begin{equation}
  C_F(t) = \langle \delta \mathbf{F}(t)\cdot\delta \mathbf{F}(0)\rangle, 
  \label{eq:15}
\end{equation}
where $\delta\mathbf{F}(t)=\mathbf{F}(t)-\langle \mathbf{F}\rangle$. 

\section{Results}
The force acting on the entire heme, $\mathbf{F}_H$, was calculated from MD simulations. This procedure averages out the short-time fluctuation of the forces caused by internal vibrations and allows us to focus on the long-time dynamics, produced by the bath, and its potential effect on the observable displacement of the iron. We found that the force-force time correlation function calculated for the iron atom is dominated by intramolecular vibration and is oscillatory (see ESI$^\dag$). The long-time dynamics is hard to extract from that correlation function, which is the reason for our focus on the overall force acting on the heme. However, this overall  force needs rescaling when applied to the individual iron atom. Assuming that the heme moves as a rigid body, the re-scaling is given by the ratio of the iron mass $m=56$ g/mol and the mass of the heme $M=614$ g/mol
\begin{equation}
\mathbf{F}= \frac{m}{M}\mathbf{F}_H .
\label{eq:16}   
\end{equation}

\begin{figure}
\includegraphics*[width=8cm]{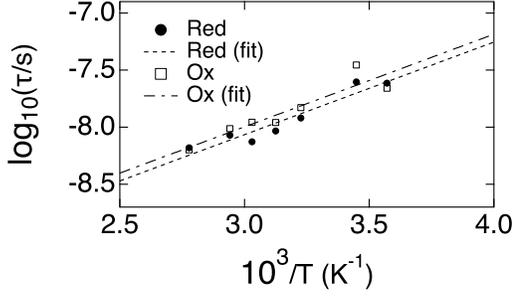}
\caption{Long relaxation time of the force-force autocorrelation function of the total force acting on the heme vs $1/T$. The results of MD simulations for the reduced (Red, filled circles) and oxidized (Ox, open squares) are fitted to Arrhenius linear functions with the slopes $E_\text{Red}/k_B=E_\text{Ox}/k_B=1868$ K. 
}
\label{fig:2}  
\end{figure}

This re-scaling, assuming the heme moving as a rigid body, can obviously apply only to the slowest dynamical components of the force. In contrast, the correlation function $C_F(t)$ calculated from simulations shows a number of time-scales, from sub-picoseconds, to long-time dynamics on the time-scale of 6--25 ns ($T\simeq 300$ K). While the slowest relaxation process usually constitutes about half of the amplitude of the time correlation function, the scaling in Eq.\ \eqref{eq:16} does not discriminate between the slow and fast dynamics. It is therefore clear that our estimate of the overall amplitude of the force acting on heme's iron is good only up to some effective coefficient accounting for imperfect rigidity of the heme. Elastic deformations of the heme shifting its center of mass are effectively disregarded in the re-scaling assuming the rigid-body motions. Given these uncertainties, we estimate $\langle \delta F^2\rangle_r$ in Eq.\ \eqref{eq:1} from the following equation
\begin{equation}
  \langle \delta F^2\rangle_r = f_\text{ne}(T) (m/M)^2 \langle \delta F_H^2\rangle .
  \label{eq:17}
\end{equation}
The nonergodicity parameter $f_\text{ne}(T)$ here comes from the dynamic restriction imposed on the integral over the frequencies in Eq.\ \eqref{eq:14}. Assuming that only the slowest component in the relaxation of the force can potentially enter the observation window, $\tau_r=142$ ns, we can write\cite{DMjpcm:15} $f_\text{ne}(T)$ in the form corresponding to exponential relaxation of $C_F(t)$ in Eq.\ \eqref{eq:15} (see below the discussion of non-exponential,  stretched dynamics)
\begin{equation}
f_\text{ne}(T) = (2/\pi) \textrm{cot}^{-1}\left[\tau(T)/\tau_r\right] .  
\label{eq:18}
\end{equation}
In this equation, $\tau(T)$  is the relaxation time of the slowest component of $C_F(t)$. A similar expression accounting for the finite resolution of the spectrometer was used in the past for the integrated elastic intensity.\cite{Springer:77}

It is clear from Eq.\ \eqref{eq:18} that the nonergodicity parameter is equal to unity when $\tau(T)\ll \tau_r$ and the fluctuations of the force are ergodic. In the opposite limit of slow fluctuations, $\tau(T)\gg \tau_r$, the force fluctuations are dynamically frozen on the observation time and do not contribute to the softening of iron's displacement, $f_\text{ne}\to 0$. This corresponds to low temperatures when intra-heme vibrations dominate. The crossover temperature $T_d$ is reached at $\tau_r\simeq\tau(T_d)$.     

The long-time relaxation times $\tau(T)$ are shown in Fig.\ \ref{fig:2}. The activation barrier of this relaxation time, $E_a/k_\text{B}\simeq 1900$ K, is below the typical values for the $\alpha$-relaxation of condensed materials, thus pointing to a localized (secondary) relaxation process of the protein-water interface.\cite{Frauenfelder:09,Khodadadi:2015jp} This relaxation time is determined in the range of temperatures $280\le T\le 360$ K, where our simulations demonstrate sufficient convergence. The Arrhenius fits of the simulation data (lines in Fig.\ \ref{fig:2}) are then extrapolated to lower temperature where the experimental M{\"o}ssbauer data are available. These extrapolated relaxation times are used in Eq.\ \eqref{eq:18} to calculate the nonergodicity factor in Eq.\ \eqref{eq:17}.       

\begin{figure}
\includegraphics*[width=8cm]{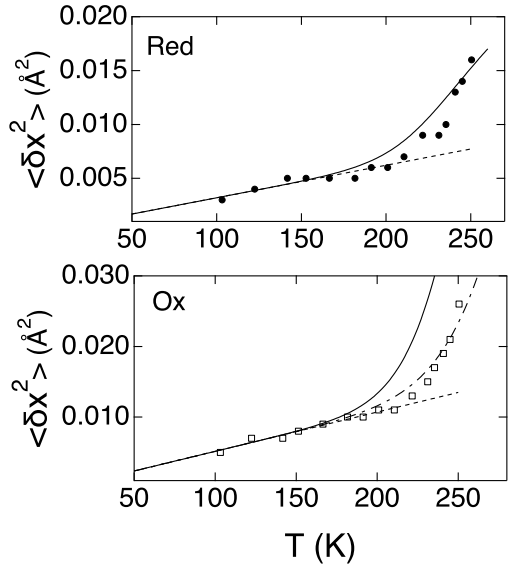}
\caption{$\langle\delta x^2\rangle$ for reduced (Red, upper panel) and oxidized (Ox, lower panel) states of Cyt-c. The points are experimental data\cite{Frolov:97} and the solid lines are calculations according to Eqs.\ \eqref{eq:1}, \eqref{eq:17}, and \eqref{eq:18}. The dashed lines are low-temperature interpolations of the experimental data. The dashed-dotted line in the lower panel is based on multiplying the relaxation time $\tau(T)$ for the Ox state with the constant coefficient equal to 2.65.  
}
\label{fig:3}  
\end{figure}

Calculations of displacements of the heme iron based on Eqs.\ \eqref{eq:1}, \eqref{eq:17}, and \eqref{eq:18} are shown in two panels of Fig.\ \ref{fig:3}. The experimental results\cite{Frolov:97} are reasonably reproduced by our calculations in the Red state of the protein without any additional fitting. The shift of the crossover temperature to a higher value in the Ox state observed experimentally would imply, in our model, slower dynamics of the force or a larger value of $\langle \delta F_H^2\rangle$. While a larger value of $\langle \delta F_H^2\rangle$ is indeed observed (Table \ref{tab:1}), its overall result is insufficient to explain the shift of the experimental crossover temperature. The experimental results are recovered by multiplying $\tau(T)$ from simulations by a factor of 2.65. While this factor is obviously arbitrary, the need for a correction might be related to our insufficient sampling of the long-time dynamics, extrapolation of the high-temperature relaxation times to lower temperatures, and the assumption of exponential dynamics not supported by measurements with protein powders\cite{Hong:2011qf,Khodadadi:2015jp,Ngai:2017kj} (see below).

\begin{table}[h]
\small
  \caption{\ Separation of $\langle \delta F_H^2\rangle$ (nN$^2$) into the electrostatic (El.) and non-polar (vdW) components and the splitting into the protein (Prot.) and water contributions ($T=320$ K). }
  \label{tab:1}
  \begin{tabular*}{0.5\textwidth}{@{\extracolsep{\fill}}lccccc}
    \hline
    Redox State & El. & vdW & Prot. & Water & Total \\
    \hline
    Red &  8.86  &  23.17 & 9.60   &   3.12  & 14.52 \\
    Ox  &  16.62 &  12.65 & 19.34  &  2.87 & 17.05\\
    
    \hline
  \end{tabular*}
\end{table}

\begin{figure}
\includegraphics*[width=8cm]{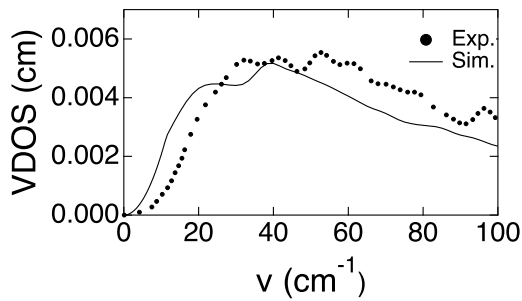}
\caption{Experimental (Exp.,\cite{Leu:08,Leu:2016fp} $T=68$ K ) and simulation (Sim., $T=300$ K) vibrational density of states for Cyt-Ox ($\bar\nu=\omega/(2\pi c)$, $c$ is the speed of light). Simulations were done for 1 ns in the NVE ensemble with non-rigid protons and 0.25 fs integration step (configurations saved every 1 fs).  
}
\label{fig:4}  
\end{figure}

Despite some difficulties with the long-time dynamics in the cyt-Ox state, the short-time dynamics produced by simulations are consistent with experiment. This is confirmed by the calculation of the vibrational density of states
\begin{equation}
D(\omega) =  \sum_{\alpha=1}^{3N} \hat (\mathbf{e}_\alpha\cdot\mathbf{\hat x}) ^2 \delta\left(\omega - \omega_\alpha \right)  , 
\label{eq:19}
\end{equation}
where $\mathbf{\hat e}_\alpha$ are expansion coefficients for the linear transformation from the Cartesian displacement of the Fe atom to normal coordinates $Q_\alpha$ in Eq.\ \eqref{eq:7}. The normalization of the density of states adopted in producing the experimental data shown in Fig.\ \ref{fig:4} requires\cite{2001JPCM...13.7707S} 
\begin{equation}
\int_{0}^\infty D(\omega) d\omega = 1   
\label{eq:20}
\end{equation}
With this normalization, the density of states from simulations was computed from the velocity-velocity autocorrelation function (see ESI$^\dag$ for more detail) and displayed in Fig.\ \ref{fig:4}. There is an excess of low-frequency modes relative to experiment,\cite{Leu:08,Leu:2016fp} which might be related to the expansion of the protein at $T=300$ K, at which simulations were performed, compared to the experimental temperature of $T=68$ K.

Table \ref{tab:1} shows the splitting of the variance of the force acting on the heme into electrostatic and van der Waals (vdW) components and, additionally, into the components from the water and protein parts of the thermal bath. Note that the components  do not add to the total force variance because of cross-correlations. The splitting into components indicates that vdW interactions and electrostatics contribute comparable magnitudes to the force variance. The softening of iron vibrations cannot therefore be fully attributed to electrostatics (dielectric effect\cite{Fenimore:02,Frauenfelder:09,Fenimore:2013eo}). It cannot be attributed to the hydration shell\cite{Fenimore:2013eo} either and is in fact a combined effect of protein and water, with the dominant contribution from the protein. The water contribution can be further diminished in solid samples used in neutron scattering or M{\"o}ssbauer spectroscopy. 

The separation of the force variance between protein and water allows us to comment on the idea of ``slaving'' of the protein dynamics by water suggested by Frauenfelder and co-workers.\cite{Fenimore:02,Fenimore:04,Frauenfelder:09} The ``slaving'' phenomenology implies the equality of the enthalpy of activation for a relaxation process in the protein with the enthalpy of activation for the structural relaxation of bulk water ($\alpha$-relaxation). When plotted in the Arrhenius coordinates ($-\ln[\tau]$ vs $1/T$) the two plots are then parallel. 

\begin{figure}
\includegraphics[width=5cm]{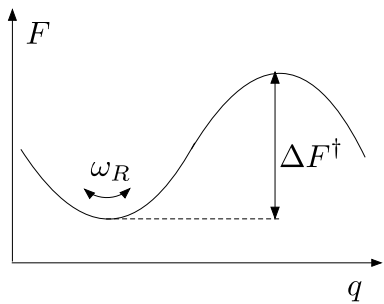}
\caption{Activated Kinetics in the Kramers' friction dominated limit. The characteristic frequency of vibrations in the well is given by $\omega_R=\omega_0^2/\zeta$ for an overdamped harmonic oscillator with the eigenfrequency $\omega_0$ and friction with the medium $\zeta$; $\Delta F^\dag$ is the free energy of activation along the reaction coordinate $q$.   
}
\label{fig:5}  
\end{figure}

The origin of this phenomenology is easy to appreciate within the framework of Kramers' activated kinetics dominated by friction with the thermal bath (Fig.\ \ref{fig:5}). The rate constant of an activated process $\propto \omega_R\exp[-\beta \Delta F^\dag]$ is the product of an effective frequency in the reactant well $\omega_R$ with the Boltzmann factor $\exp[-\beta \Delta F^\dag]$ involving the free energy of activation $\Delta F^\dag$. If the motions along the reaction coordinate are represented by an overdamped harmonic oscillator with the frequency $\omega_0$ and the friction coefficient $\zeta$, the direct solution of the Langevin equation leads to the relaxation frequency $\omega_R=\omega_0^2/\zeta$. Therefore, ``slaving'' appears when most of energy dissipation occurs to the water part of the thermal bath (which has a higher heat capacity than the protein\cite{Cooper:1984tv}). In that case, the temperature dependence of $\zeta(T)$, and of the corresponding relaxation process in water,  would determine the temperature dependence of the relaxation rate in the protein, which is only shifted to lower rates due to an additional activation barrier $\Delta F^\dag$ (assuming $\Delta F^\dag$ is temperature-independent). This is the ``slaving'' scenario.       

While there are reported instances when this picture is correct,\cite{Khodadadi:2015jp,Khodadadi:2017dg} one can argue that energy dissipation for localized processes occurs to the protein hydration shell, which possesses its own relaxation spectrum. Indeed, Frauenfelder and co-workers\cite{Frauenfelder:09} argued that localized processes in the protein have to be ``slaved'' to the relaxation of the hydration layer. Consistently with that notion, relaxation processes related to protein function are often characterized by the activation barrier much lower than those for $\alpha$-relaxation of bulk water (Fig.\ \ref{fig:2}). For instance, the Stokes shift dynamics directly related to the redox activity of Cyt-c show the activation barrier of its relaxation time $E_a/k_\text{B}\simeq 840$ K.\cite{DMjpcb2:17} This is much lower than $\sim 1560$ K (increasing to $\sim 6400$ K upon cooling) from diffusivity and viscosity of water ($\alpha$-relaxation).\cite{Dehaoui:2015ii} The idea of ``slaving'' to $\beta$-relaxation of the hydration shell is less useful, and is harder to prove, since  relaxation of the shell is mostly inaccessible experimentally. Our simulation results allow us such a test since the dynamics of both the hydration layer and of the heme's iron are available. 

\begin{figure}
\includegraphics[width=8cm]{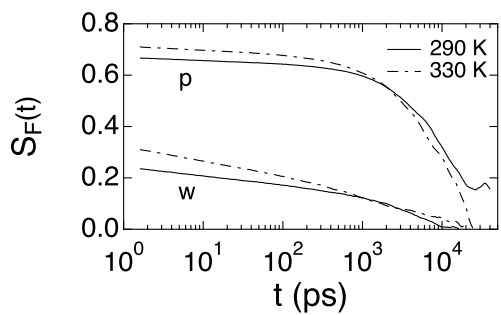}
\caption{Normalized force-force correlation function $S_F(t)=C_F(t)/C_F(0)$ for the protein (p) and water (w) components at the temperatures indicated in the plot. 
}
\label{fig:6}  
\end{figure}

\begin{figure}
\includegraphics[width=8cm]{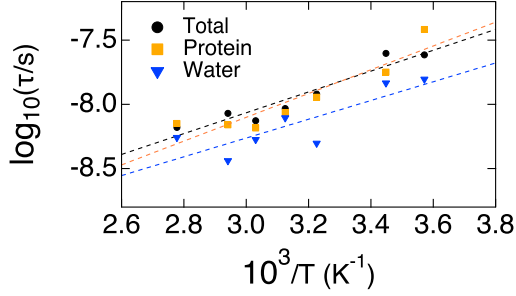}
\caption{Long relaxation time of the force-force autocorrelation function of the force acting on the heme vs $1/T$ (black circles) for reduced Cyt-c. Also shown are the relaxation times for the force on the heme produced by the protein (squares) and by water (triangles). Fits to Arrhenius linear functions are shown by the dashed lines.  
}
\label{fig:7}
\end{figure}

In application to M{\"o}ssbauer experiment, our data do not support ``slaving''.  Only $\sim 20$\% of the force variance acting on heme's iron comes from from hydration water (Table \ref{tab:1}). This also implies that the dynamics should be biomolecule-specific.\cite{Khodadadi:2010qf} In this scenario, ``slaving'' would be only possible if the protein dynamics followed the dynamics of water. The results of simulations do not support this conjecture: the dynamics of $S_F(t)=C_F(t)/C_F(0)$ are distinctly different for the protein and its hydration water (Fig.\ \ref{fig:6}). The dynamics of water is on average significantly faster (a larger drop from the initial value $S_F(0)=1$, not resolved in Fig.\ \ref{fig:6}), and the slow dynamics of the protein and water are not consistent either.  Nevertheless, the temperature dependence of the relaxation time of the force-force correlation function is consistent between the protein and water components (Fig.\ \ref{fig:7}). The enthalpies of activation for the protein and water relaxation are, therefore, close in magnitude, in a general accord with the ``slaving'' phenomenology. The origin of this effect can be traced to coupled fluctuations of the protein and hydration water,\cite{LiT:07,Halle:09,Furse:2010dw,DMjpcb1:12,ContiNibali:2014ii} without invoking a dominant role of water in the  dynamics.

\begin{figure}
\includegraphics[width=8cm]{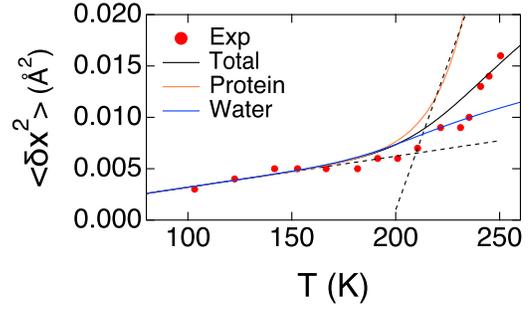}
\caption{$\langle\delta x^2\rangle$ for the reduced state of Cyt-c. The points are experimental data\cite{Frolov:97} and the solid lines are calculations according to Eqs.\ \eqref{eq:1}, \eqref{eq:17}, and \eqref{eq:18}. The calculations are done for the total force-force correlation function (black) and for its components from the protein (orange) and water (blue). The dashed lines refer to the low-temperature linear fit of the experimental data and to the high-temperature linear fit of the iron displacement produced by the protein. }
\label{fig:8}
\end{figure}

Water is a faster subsystem producing a shorter relaxation time of $C_F(t)$. One therefore anticipates that the temperatures of ergodicity breaking should separate for the water and protein components of the thermal bath.\cite{2017arXiv170503128B} This indeed happens, as is illustrated in Fig.\ \ref{fig:8} for the reduced state of Cyt-c. The rise of $\langle \delta x^2\rangle$ due to water occurs at $\simeq 150$ K, while the transition temperature for the protein is $\simeq 200$ K. The water's onset is hard to disentangle because the force produced by water on the heme is relatively low. One might expect that the water transition is better resolved in neutron scattering experiments\cite{2017arXiv170503128B} since a large number of protons located close to the interface potentially contribute to the signal. Overall, this calculation clearly points to a nonergodic origin of the dynamical transition, as we stress again below when considering the separation of rotational and translational motions of water in the hydration shell.

\section{Stretched relaxation} 
Difficulties with reproducing ergodicity breaking of Cyt-Ox (Fig.\ \ref{fig:3} lower panel) might be related to a limited applicability of the results obtained for solutions to dynamics in protein powders and crystals studied experimentally. In addition to the obvious uncertainty of extrapolating the high-temperature simulation results to lower temperatures, the dynamics of hydration water can be qualitatively different in those environments compared to solutions. The relaxation of hydration water in powders was associated by Ngai and co-workers\cite{Ngai:2013ic,Ngai:2017kj} with the general phenomenology of confined water in water-containing glass-formers. The $\nu$-process characterizing such dynamics is highly stretched, with a very slow decay of the high frequency tail of the loss function: $\epsilon''\propto \omega^{-\gamma}$ for the dielectric loss\cite{Nickels:2013jw,Nakanishi:2014gy} and $\chi''(\omega)\propto \omega^{-\gamma}$  for the neutron scattering loss.\cite{Hong:2011qf} A low value of stretching exponent, $\gamma\simeq 0.2$, is observed in both cases.

\begin{figure}
\includegraphics*[width=8cm]{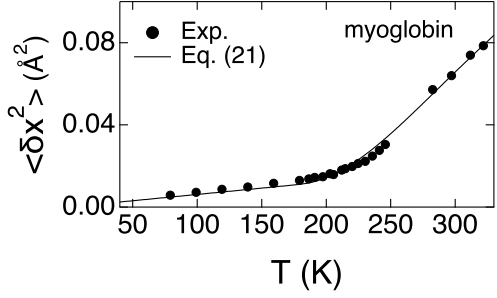}
\caption{MSF of heme iron in oxidized myoglobin. Points indicate experimental results,\cite{Parak:05} solid line refers to the fit to Eq.\ \eqref{eq:1} and \eqref{eq:17} with the nonergodicity factor $f_\text{ne}(T)$ determined from stretched dynamics according to Eq.\ \eqref{eq:21}. The nonergodic force variance is determined according to Eq.\ \eqref{eq:22} with the fitting constant $A=2.5$ nN/\AA\ (corresponds to $\langle \delta F^2\rangle = 0.1$ nN$^2$ at $T=300$ K).  
}
\label{fig:9}
\end{figure}

The $\nu$-process observed in lysozyme and myoglobin powders by dielectric spectroscopy was identified to cause the dynamical transition in neutron scattering.\cite{Khodadadi:2017dg} We can therefore use the corresponding relaxation time $\tau(T)$ reported from dielectric measurements to explain M{\"o}ssbauer data for met-myoglobin\cite{Parak:05} (oxidized form of myoglobin). Before we do that, we have to extend the nonergodicity parameter obtained in Eq.\ \eqref{eq:18} for exponential relaxation to stretched exponential relaxation. Cole-Cole function was used to fit the dielectric data.\cite{Nakanishi:2014gy} We therefore can re-write the nonergodicity parameter $f_\text{ne}(T)$ as follows
\begin{equation}
  f_\text{ne}(T) = \frac{2}{\pi} \int_{\tau(T)/\tau_r}^{\infty} \frac{d\omega}{\omega} \mathrm{Im}\left[(1+(i\omega)^\gamma)^{-1} \right] , 
  \label{eq:21}
\end{equation}
where  $\gamma$ is the stretching exponent of the Cole-Cole function. At $\gamma=1$, Eq.\ \eqref{eq:21} transforms to Eq.\ \eqref{eq:18}. This nonergodicity factor can be used in the following form for the force variance 
\begin{equation}
\beta\langle \delta F^2\rangle_r = A  f_\text{ne}(T)
\label{eq:22}  
\end{equation}
where, according to the standard prescription of the fluctuation-dissipation theorem, the amplitude $A$ is held constant. The use of this form along with $\gamma=0.25$ and the experimental $\tau(T)$ (see Fig.\ S4 in ESI$^\dag$) in Eq.\ \eqref{eq:21} produce the MSF of myoglobin shown by the solid line in Fig.\ \ref{fig:9}. The fit requires $\langle \delta F^2\rangle\simeq 0.1$ nN$^2$ at $T=300$ K, which is roughly consistent with $\langle \delta F^2\rangle \simeq 0.14$ nN$^2$ for Cyt-Ox in Table \ref{tab:1} when Eq.\ \eqref{eq:16} is applied. The quality of the fit is significantly reduced with $\gamma=1$, which testifies to the need of applying stretched relaxation to describe ergodicity breaking in protein powders.

\section{Glass transition}
The lower crossover temperature $T_g$ of the protein MSF represents the glass transition of the hydration shell. It was previously identified with the onset of translational diffusion of the water molecules in the shell.\cite{Zanotti:1999jl} However, glass science requires one to pay attention not only to translations, but also to molecular rotations. There are a number of reasons for that. First, the configurational entropy of fragile glass-formers is mostly rotational\cite{DMjcp4:16} (e.g., the heat capacities of supercooled ethanol and its plastic crystal are nearly identical\cite{Kabtoul:2008ie}). Reducing the configurational entropy is required for reaching the glass transition\cite{Richert:98} and, therefore, the rotational configuration space has to be strongly constrained close to $T_g$. Second, the temperature dependence of the dielectric relaxation time can be superimposed with the relaxation time from viscosity\cite{Ediger:96} and with the diffusion coefficient. Therefore, both rotations and translations are expected to dynamically freeze near $T_g$. 

The density of water in the hydration shell is enhanced compared to the bulk,\cite{Gerstein:1996tg,Svergun:98} and shell water, being heterogeneous and more disordered than the bulk,\cite{Lerbret:2012gz,ContiNibali:2014ii} is close in physical properties to a mixture of low-density and high-density amorphous ice.\cite{Paciaroni:2008ge} Nevertheless, the positional structure of the shell (pair distribution function) does not change with cooling, and there is no structural transition associated with crossing the temperature $T_d$.\cite{Yoshida:2016il}  Compared to the positional structure and diffusional dynamics,\cite{Makarov:00} there is much less experimental and computational evidence on orientational correlations and rotational dynamics of water in the hydration shell. The single-particle rotational dynamics are slowed down by a factor of 2--4, as is seen by NMR\cite{Halle:04} and computer simulations.\cite{Laage:2017ig}  Collective relaxation probed by Stokes shift of optical dyes are much slower, in the range of sub- to nanonoseconds,\cite{Jordanides:99,Pal:04,Abbyad:07} pointing to a significantly slower collective response of water dipoles\cite{DMjpcl:15} compared to single-molecule rotations.         

\begin{figure}
\includegraphics*[width=8cm]{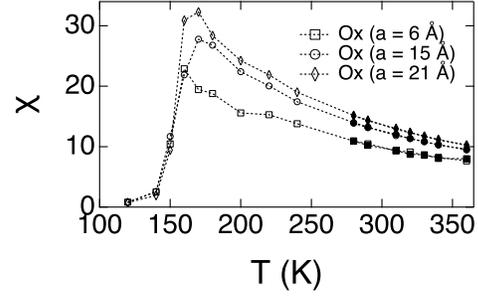}
\caption{The dipolar susceptibility of the hydration shell water calculated from MD simulations according to Eq.\ \eqref{eq:23} for shells of thickness $a$ around Cyt-Ox (open points) and Cyt-Red (filled points) at different temperatures (some Red and Ox points coincide on the scale of the plot). The dotted lines connect the points to guide the eye.  
}
\label{fig:10}
\end{figure}

The fact that the collective response of the shell dipole is quite different from single-particle MSF is illustrated in Fig.\ \ref{fig:10}, which shows the dipole moment variance for hydration shells of Cyt-Ox and Cyt-Red with varying temperature and thickness of the shell. More specifically, we present the dimensionless variance of the shell dipole moment defined analogously to the dielectric susceptibility of bulk dielectrics 
\begin{equation}
\chi(a) = [3k_\text{B}Tv_w N_w(a)]^{-1} \langle \delta\mathbf{M}(a)^2\rangle .
  \label{eq:23}
\end{equation}
Here, $v_w$ is the volume of a single water molecule (effective diameter $2.87$ \AA\cite{DMjpc:95}) and $N_w(a)$ is the number of water molecules in the shell of thickness $a$ measured from the van der Waals surface of the protein; $\mathbf{M}(a)$ is the total dipole moment of the water molecules in the shell, $\delta\mathbf{M}(a)=\mathbf{M}(a)-\langle \mathbf{M}(a)\rangle$.  

The main qualitative difference between the temperature dependence of the MSF and the shell dipole is that the latter clearly violates the fluctuation dissipation theorem,\cite{Kubo:66} which predicts $ \langle \delta\mathbf{M}(a)^2\rangle\propto T$. The phenomenology of susceptibility decaying with temperature, in violation of the fluctuation-dissipation theorem,\cite{DMjcp1:16} is shared by most polar liquids.\cite{Richert:2014wa} However, in contrast to homogeneous liquids, the protein hydration shells are heterogeneous and highly frustrated.\cite{DMjpcl:15}  This is because polarized interfacial water has to follow a nearly uniform mosaic of positively and negatively charged surface residues. Surface charges orient water dipoles into polarized domains. These domains are mutually frustrated by altering sign of the charged residue, but stay in the fluid state with the fluctuations of the shell dipole significantly slowed down (hundreds of picoseconds to nanoseconds\cite{DMjpcl2:12,DMjpcl:15}) compared to the bulk. This new physics, quite distinct from bulk polar liquids, connects hydration shells to relaxor ferroelectrics, where mutual frustration of dipolar crystalline cells breaks the material into ferroelectric nanodomains at the glass transition reached above the Curie point.\cite{Samara:03} 

The phenomenology of relaxor ferroelectrics suggests that the dipolar response of the shell is determined by reorienting the polarized domains, instead of predominantly single-particle rotations found in bulk polar liquids.\cite{Madden:84} This interpretation is supported by nanosecond time-scales characterizing the dynamic susceptibility of the shell $\chi(\omega,a)$\cite{DMjpcl:15} (see the ESI$^\dag$). This picture does not contradict to the dynamic (fluid) nature of the hydration shell in which water can diffuse along the surface visiting a residue per $\simeq 11$ ps.\cite{Hospital:2017eh} Moving from a positive to a negative residue can be accompanied with a dipole flip, still preserving the domain structure, which requires much longer times to be altered. The dipole flip of a water molecule moving to a neighboring residue will also produce a short relaxation time for single-particle rotations.\cite{Halle:04,Laage:2017ig}

\begin{figure}
\includegraphics*[width=8cm]{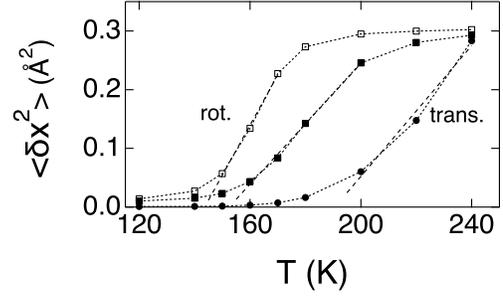}
\caption{Center of mass MSF (trans., circles) and the MSF due to molecular rotations (rot., squares) of water molecules within the hydration shell 6 \AA\ thick around the Ox Cyt-c. The center of mass translations and molecular rotations are calculated within the time-window of 100 ps (filled points) and 1 ns (open points). The MSFs for center-of-mass translations are reduced by a factor of 40 to bring them to the same scale with the results for rotations. The dashed lines are linear fits through subsets of points to illustrate differences in the onset temperatures ($T_\text{rot}(\mathrm{1\ ns})=144$ K, $T_\text{rot}(\mathrm{100\ ps})=152$ K, and $T_\text{tr}(\mathrm{100\ ps})=191$ K. The dotted lines connecting the points are drawn to guide the eye.  
}
\label{fig:11}
\end{figure}

A sharp drop of $\chi(a)$ at about $\simeq 145$ K signals reaching the glass transition on the time scale of MD simulations (Fig.\ \ref{fig:10}). This $T_g$ is somewhat lower than experimental $T_g\simeq 170$ K from calorimetry of concentrated solutions of Cyt-c.\cite{Green:94} The glass transition of the hydration shell prevents elastic motions of the protein, making a hydrated protein harder at low temperatures than the dry one.\cite{Hong:2013bk} One wonders if rotations and translations of water molecules in the shell terminate at the same temperature. Figure \ref{fig:11} shows that this is not the case (see ESI$^\dag$ for the details of calculations). The glass transition for $\chi(a)$ coincides with freezing of water rotations. The onset temperature depends on the observation window (cf.\ filled to open squares in Fig.\ \ref{fig:11}), consistent with ergodicity breaking at the transition. On the contrary, the onset of water translations occurs at a higher temperature, $\simeq 190$ K. A similar phenomenology was recently reported from neutron scattering of protein's hydration shell,\cite{Schiro:2015cv} where the onset of water's translations also followed the onset of rotations.  The temperature of translational onset is close to $T_d$, as was noted in the past.\cite{Tarek_PhysRevLett:02}        

A crude estimate of the ``dielectric constant'' of the shell might be relevant here. If, for the sake of an estimate, one adopts the connection between the dielectric constant and the susceptibility of bulk dielectrics, $\epsilon(a)=1+4\pi\chi(a)$, then the inspection of Fig.\ \ref{fig:10} suggests $\epsilon(a)\simeq 407$ at $T=170$ K and $a=21$ \AA. This very high dielectric constant is consistent with recent dielectric spectroscopy of protein powders,\cite{Shinyashiki:2009jd,Nakanishi:2014gy} reporting high dielectric increments $\Delta \epsilon\simeq 10^2-10^4$ for the relaxation process reaching 1-10 $\mu$s at the room temperature. Given the temperature dependence of this relaxation process, it appears likely that it is responsible for glass transition of hydrated protein samples.\cite{Khodadadi:2015jp} The drop of $\chi$ at $T_g$ seen in Fig.\ \ref{fig:10}, and a similar behavior observed previously in simulations of lysozyme,\cite{DMjpcl:15} suggests a possible connection between high $\Delta \epsilon$ and polarized domains formed in the hydration shell.

\section{Onset of protein functionality}
Equation \eqref{eq:1} offers a natural explanation of the extended flexibility of proteins at high temperatures in terms of the force constant assigned to a cofactor or residue in the folded protein.\cite{Hong:2013bk} According to Eq.\ \eqref{eq:1}, softening of the protein matrix due to collective agitation of the protein-water thermal bath reduces the vibrational force constant $\kappa_\text{vib}=(\beta\langle \delta x^2\rangle_\text{vib})^{-1}$ by the magnitude 
\begin{equation}
\kappa_b=\beta \langle \delta F^2\rangle_r,  
\label{eq:24}
\end{equation}
which reduces the total force constant $\kappa= (\beta\langle \delta x^2\rangle)^{-1}$  
\begin{equation}
\kappa = \kappa_\text{vib}-\kappa_b .  
\label{eq:25}  
\end{equation}

\begin{figure}
\includegraphics*[width=8cm]{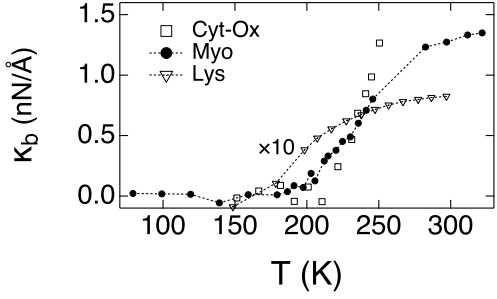}
\caption{Force constant of the protein-water medium $\kappa_b=\beta\langle \delta F^2\rangle_r$ calculated from $\kappa(T)$ and $\kappa_\text{vib}(T)$ according to Eq.\ \eqref{eq:25}. Points indicate the experimental results for Cyt-c (Ox),\cite{Frolov:97} myoglobin (Myo),\cite{Parak:05} and for lysozyme dissolved in 50:50 glycerol-D$_2$O solvent at $h=0.83$ g D$_2$O/g Lys.\cite{Paciaroni:02}  The results for lysozyme are multiplied by a factor of 10 to bring them to the scale of the plot. The dotted lines connecting the points are drawn to guide the eye. 
}
\label{fig:12}
\end{figure}

Using  Eq.\ \eqref{eq:25},  Fig.\ \ref{fig:12} shows $\kappa_b(T)$  for Cyt-c (Ox) and myoglobin (Figs.\ \ref{fig:3} and \ref{fig:9}). We have additionally included the results from neutron scattering of lysozyme (Lys) in 50:50 glycerol-D$_2$O solution ($h=0.83$ g D$_2$O/g Lys),\cite{Paciaroni:02} which display a crossover temperature at $\simeq 180$ K. All these data point to a rise of $\kappa_b$ at $T_d$ to a nearly constant value charactering the protein flexibility at GHz frequencies. The Young's moduli of the hydrated protein fall with increasing temperature\cite{Perticaroli:2013bn,Hong:2013bk} in a fashion consistent with $\kappa_b$ in Fig.\ \ref{fig:12}.   

The notion of protein dynamics as proxy for enzymatic activity has been actively discussed in the recent literature.\cite{Henzler-Wildman:2007ly} One has to clearly distinguish flexibility,\cite{Cooper:1984tv} i.e. the ability to sample a large number of conformations, from the actual dynamics, i.e. the time-scales involved in usually dissipative decay of correlation functions. Whether flexibility and activity must accompany each other for slow (in milliseconds) enzymetic reactions remains to be seen,\cite{Kamerlin:10} but there is one class of enzyme reactions where protein configurational space has to be dynamically restricted for the reaction to occur.\cite{DMjpcm:15} This is the process of protein electron transport essential to production of all energy in biology, either through photosynthesis or through mitochondrial respiration.\cite{Bioenergetics3} 

The fluctuation-dissipation theorem connects fluctuations to response to an external perturbation.\cite{Kubo:66} In this framework, high flexibility implies high solvation,\cite{Hummer:98} or trapping, energy. Electrons in biological energy chains have to perform many tunneling steps within a narrow energy window consistent with the energy input from food or light. In order to accomplish vectorial electron transport, energy chains have to avoid deep energy traps. Therefore, large conformational motions producing asymmetries in solvation energies between the initial and final tunneling states have to be dynamically frozen on the reaction time.\cite{DMjpcm:15} 

This phenomenology is consistent with what we have found here for the dynamical transition of atomic displacements. The role of the force constant in Eq.\ \eqref{eq:24} is played by the reorganization energy $\lambda$ determined through the variance of the donor-acceptor energy gap $X$ used to gauge the progress of the reaction. The reorganization energy is determined through the variance of $X$ by the equation inspired by the fluctuation-dissipation theorem (cf.\ to Eq.\ \eqref{eq:24})
\begin{equation}
\lambda(k_R) = \beta \langle \delta X^2\rangle_r/2 
\label{eq:26}  
\end{equation}
Here, $\langle \delta X^2\rangle_r = \langle \delta X^2\rangle f_\text{ne}(T)$ depends on the observation window through the nonergodicity factor $f_\text{ne}(T)$ (Eq.\ \eqref{eq:18}) multiplying the thermodynamic ($\tau_r\to\infty$) variance $\langle \delta X^2\rangle$. The only difference of this problem from our discussion of iron's MSF is that one has to replace the relaxation time of the force $\tau(T)$ with the relaxation time $\tau_X(T)$ of the Stokes-shift correlation function $C_X(t)=\langle\delta X(t)\delta X(0)\rangle$. The role of the observation window is now played by the reaction time $\tau_r=k_R^{-1}$ given in terms of the reaction rate constant $k_R$.         

\begin{figure}
\includegraphics*[width=8cm]{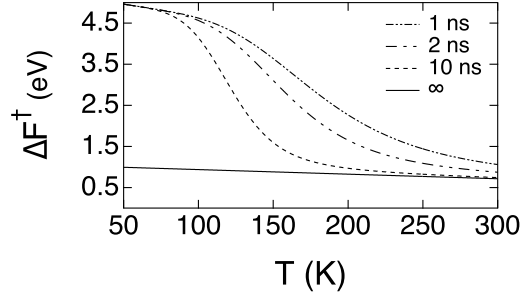}
\caption{$\Delta F^\dag$ given by Eq.\ \eqref{eq:26} vs $T$  calculated from MD simulations ($\sim 250$ ns of simulations at each temperature\cite{DMjpcb2:17}). The legend indicate the reaction times $\tau_r = k_R^{-1}$. Deviations from the thermodynamic behavior, $k_R=0$, are determined by the nonergodic factor $f_\text{ne}(T)$ (Eq.\ \eqref{eq:18}) calculated from the observation window $\tau_r=k_R^{-1}$ and the relaxation time\cite{DMjpcb2:17} $\tau_X(T)(\mathrm{s})=\exp[-23.8 + 835/T]$. The reorganization energies from long simulation trajectories are approximated by linear functions of temperature: $\lambda^\text{St}(T)= 1.71 - 0.0015\times T$ eV, $\lambda(T)=4.19 - 0.00446\times T$ eV ($T$ is in K).     
}
\label{fig:13}
\end{figure}
      
The reorganization energy $\lambda(k_R)$ quantifies the depth of the trap created for a charge by the protein-water thermal bath. The amount of energy to de-trap the electron and bring it back to the tunneling configuration specifies the activation barrier $\Delta F^\dag$. It is given in terms of two energy parameters:\cite{DMjpcm:15} the difference of first moments of $X$ in the initial and final states, known as the Stokes-shift reorganization energy $\lambda^\text{St}$, and the second moment of $X$ specified by $\lambda(k_R)$
\begin{equation}
\Delta F^\dag = (\lambda^\text{St})^2/[4\lambda(k_R)] . 
\label{eq:26} 
\end{equation}

The parameter $\lambda^\text{St}$ specifies the energy difference between two states of the protein (Red and Ox in the case of Cyt-c). It does not reach its thermodynamic value  because of the inability of the protein to sample its entire phase space on the reaction time.\cite{DMjpcm:15} Instead of reaching, through a conformational change, two thermodynamic minima of stability (for Red and Ox states), the protein gets trapped in intermediate local minima. The time separation $k_R^{-1} \ll \tau_\text{conf}$ between the reaction time and the time of the conformational transition $\tau_\text{conf}$ constrains the availvble configuration space allowing a relatively small value of $\lambda^\text{St}$ such that the condition $\lambda^\text{St}\ll\lambda(k_R)$ keeps the reaction barrier in Eq.\ \eqref{eq:26} relatively low. The reorganization energy $\lambda(k_R)$ in the denominator in Eq.\ \eqref{eq:26} is, however, directly affected by nonergodic freezing of a subset of degrees of freedom, which can lead to a significant increase of the reaction barrier at low temperatures and to the termination of the protein function. 

This perspective is illustrated in Fig.\ \ref{fig:13} showing the effect of the observation window on $\Delta F^\dag(T)$. The input parameters to the results shown in Fig.\ \ref{fig:13} are $\lambda^\text{St}(T)$ and $\lambda(T)$ taken from long trajectories ($k_R\to 0$) and the Stokes-shift relaxation time $\tau_X(T)$ calculated for Cyt-c.\cite{DMjpcb2:17}  As the temperature decreases, the relaxation time $\tau_X(T)$ leaves the observation window, $\tau_r=k_R^{-1}$, and $\lambda(k_R)$ drops. The activation barrier grows at low temperatures (see Eq.\ \eqref{eq:26}) and the reaction slows down due to ergodicity breaking qualitatively consistent with the dynamical transition for the atomic MSF (at faster rates, such as those involved in primary events of photosynthesis, $\lambda^\text{St}$ becomes affected by $k_R$ and the picture changes again\cite{DMpccp:10}). The overlap of the time-scales probed by the neutron scattering and M{\"o}ssbauer spectroscopy with the typical reaction times of protein electron transfer suggests that the fluctuations of the protein-water thermal bath responsible for the high-temperature part of the displacement curve are the same as those involved in activating redox activity of proteins.

\section{Conclusions}
The present model assigns atomic displacements in the protein to two factors: (i) high-frequency vibrations within the subunit (residue, cofactor, etc.) and (ii) fluctuations in the position of the subunit caused by thermal fluctuations of the entire protein and its hydration shell. The second component enters the observable MSF in terms of the variance of the force applied to the center of mass of the subunit (denominator in Eq.\ \eqref{eq:1}). This equation can be alternatively viewed as softening of a stiff vibrational force constant by the protein-water thermal bath (Eq.\ \eqref{eq:25}). Since the variance of the force depends on the observation window, softening of vibrations is achieved at the temperature above $T_d$ allowing the long-time relaxation of the force autocorrelation function to remain within the observation window. An experimental link to this picture is provided by inelastic x-ray scattering\cite{Liu:2008gs,Wang:2014je,Shrestha:2017ku} recording softening of the protein phonon-like modes representing global vibrations. In line with the common observations of the dynamical transition, softening of the protein phonon modes is strongly suppressed in dry samples.\cite{Wang:2014je} Similar phenomenology is provided by the temperature dependence of the protein boson peak\cite{Perticaroli:2013bn,Frontzek:2014ie}  reflecting the density of protein collective vibrations on the length-scale of a few nanometers and THz frequency.\cite{Duval:90,Tarek:2001hw,Schiro:2011fk} For instance, the frequency of the boson peak for myoglobin falls from $\sim 32$ cm$^{-1}$ to $\sim 16$ cm$^{-1}$ when the temperature is raised from 170 to 295 K.\cite{Perticaroli:2013bn} 

The forces produced by the protein-water thermal bath at internal sites inside the protein are strongly affected by the structure and dynamics of the hydration shell.\cite{DMpre:11,Hong:2012dsa,Wood:2012bf} Shell dipoles cluster in nanodomains pinned by charged surface residues. Dynamical freezing of these nanodomains occurs at the glass transition of the hydration shell corresponding to the lower crossover temperature $T_g$ (Fig.\ \ref{fig:1}). Rotations of water molecules in the shell dynamically freeze at this temperature. Translations dynamically freeze at a higher temperature close to $T_d$. Therefore, the existence of two crossover temperatures in the dynamical transition of proteins reflects two separate ergodicity breaking crossovers for rotations and translations of hydration water (Fig.\ \ref{fig:11}).   

The entrance of the relaxation time into the resolution window, resulting in the dynamical transition of a specific relaxation mode, is often considered to be a ``trivial'' effect, in contrast to an anticipated true structural transition.\cite{DosterNature:89,Cupane:2014hka} However, this ergodicity breaking allows protein-driven reactions to proceed without being trapped into deep solvation wells. The link between flexibility and solvation, and thus the ability to produce traps, has been under-appreciated in the literature on enzymatic activity. As an illuminating example, protein electron transfer occurs in dynamically quenched proteins where ergodicity breaking prevents from developing deep solvation traps along the electron-transport chain.

\textbf{Conflict of Interests}. 
There are no conflicts of interest to declare

\textbf{Acknowledgement}. 
This research was supported by the NSF (CHE-1464810) and through XSEDE (TG-MCB080116N). We are grateful to Antonio Benedetto for useful discussions and the preprint (Ref.\ \cite{2017arXiv170503128B}) made available to us.  We acknowledge help by Daniel Martin with the analysis of the simulation trajectories.

%%%END OF MAIN TEXT%%%


%\balance


%%%REFERENCES%%%

%\bibliography{dielectric,dm,statmech,glass,protein,liquids,solvation,dynamics,simulations,surface,water,nano,lcold,ferro,bioet}
%\bibliographystyle{naturemag}

\end{document}